\documentclass[twocolumn,apjl]{aastex62}
\usepackage{ifthen}
\usepackage{etoolbox}

\usepackage{import}
\usepackage{longtable}
\usepackage{graphicx}
\usepackage{amsmath,amssymb}
\usepackage{color}
\usepackage{units}
\usepackage{epstopdf}
\usepackage{hyperref}
\usepackage{multirow}
\usepackage[caption=false]{subfig}
\usepackage{float}

\usepackage{import}
\usepackage{longtable}
\usepackage{graphicx}
\usepackage{amsmath,amssymb}
\usepackage{color}
\usepackage{units}
\usepackage{epstopdf}
\usepackage{hyperref}
\usepackage{multirow}
\usepackage[caption=false]{subfig}
\usepackage{float}

\newcommand{\beq}{\begin{equation}}
\newcommand{\eeq}{\end{equation}}
\newcommand{\bdm}{\begin{displaymath}}
\newcommand{\edm}{\end{displaymath}}

\definecolor{Gray}{gray}{0.9}
\definecolor{orange}{rgb}{0.9,0.5,0}

\newcommand*\diff{\mathop{}\!\mathrm{d}}

\newcommand{\dist}{{\mathcal D}}

\newcommand{\icecube}[0]{IceCube}

\def\newacronym#1#2#3{\gdef#1{#3 (#2)\gdef#1{#2}}}

\newacronym{\gt}{Georgia Tech}{Georgia Institute of Technology}
\newacronym{\icecube}{IceCube}{IceCube South Pole Neutrino Observatory}
\newacronym{\cta}{CTA}{Cherenkov Telescope Array}
\newacronym{\sdsc}{SDSC}{San Diego Supercomputer Center}
\newacronym{\cra}{CRA}{Center for Relativistic Astrophysics}
\newacronym{\nr}{NR}{numerical relativity}
\newacronym{\ornl}{ORNL}{Oak Ridge National Laboratory}
\newacronym{\lisa}{LISA}{Laser Interferometer Space Antenna}
\newacronym{\hst}{HST}{Hubble Space Telescope}
\newacronym{\jwst}{JWST}{James Webb Space Telescope}
\newacronym{\alma}{ALMA}{Atacama Large Millimeter Array}
\newacronym{\ligo}{LIGO}{Laser Interferometer Gravitational Wave Observatory}
\newacronym{\sph}{SPH}{smooth particle hydrodynamics}
\newacronym{\tsi}{TSI}{Terascale Supernova Initiative}
\newacronym{\wmap}{WMAP}{the Wilkinson Microwave Anisotropy Probe}
\newacronym{\cmb}{CMB}{cosmic microwave background}
\newacronym{\ibbh}{IBBH}{intermediate binary black hole}
\newacronym{\grb}{GRB}{Gamma-ray burst}
\newacronym{\grbs}{GRBs}{Gamma-ray bursts}
\newacronym{\hpc}{HPC}{High Performance Computing}
\newacronym{\htc}{HTC}{High Throughput Computing}
\newacronym{\bssn}{BSSN}{Baumgarte-Shapiro-Shibata-Nakamura}
\newacronym{\amr}{AMR}{adaptive mesh refinement}
\newacronym{\mpi}{MPI}{Message Passing Interface}
\newacronym{\cuda}{CUDA}{Compute Unified Device Architecture}
\newacronym{\gui}{GUI}{Graphical User Inferface}
\newacronym{\oit}{OIT}{Office of Information Technologies}
\newacronym{\lbg}{LBGs}{Lyman break galaxies}
\newacronym{\mhd}{MHD}{magneto-hydrodynamics}
\newacronym{\imf}{IMF}{initial mass function}
\newacronym{\sms}{SMS}{supermassive star}
\newacronym{\posix}{POSIX}{Portable Operating System Interface}
\newacronym{\ssd}{SSD}{Solid State Drive}
\newacronym{\os}{OS}{Operating System}
\newacronym{\pace}{PACE}{Partnership for an Advanced Computing Environment}
\newacronym{\oit}{OIT}{Office of Information Technology}
\newacronym{\gtswd}{GTSWD}{Georgia Tech Software Distribution}
\newacronym{\uga}{UGA}{University of Georgia}
\newacronym{\mahpic}{MaHPIC}{Malaria Host-Pathogen Interaction Center}
\newacronym{\nih}{NIH}{National Institutes of Health}
\newacronym{\cdc}{CDC}{Centers for Disease Control and Prevention}
\newacronym{\xsede}{XSEDE}{Extreme Science and Engineering Discovery Environment}
\newacronym{\osg}{OSG}{Open Science Grid}
\newacronym{\egi}{EGI}{European Grid Infrastructure}
\newacronym{\vo} {VO}{Virtual Organization}
\newacronym{\sox}{SoX}{Southern Crossroads}
\newacronym{\slr}{SLR}{Southern Light Rail}
\newacronym{\ddn}{DDN}{DataDirect Networks}
\newacronym{\rsv}{RSV}{Resource and Service Validation}
\newacronym{\cvmfs}{CVMFS}{CernVM File System}
\newacronym{\nfs}{NFS}{Network File Sytem}
\newacronym{\lsc}{LSC}{LIGO Scientific Collaboration}

\newacronym{\DA}{DA}{data analysis}
\newacronym{\CBC}{CBC}{Compact Binary Coalescences}
\newacronym{\lvc}{LVC}{LIGO-Virgo Collaboration}
\newacronym{\aligo}{aLIGO}{Advanced LIGO}
\newacronym{\ldr}{LDR}{LIGO Data Replicator}

\def\si#1{Senior Investigator#1 (SI#1)\gdef\si{SI}}
\def\emri#1{Extreme Mass-Ratio Inspiral#1 (EMRI#1)\gdef\emri{EMRI}}
\def\imbh#1{intermediate mass black hole#1 (IMBH#1)\gdef\imbh{IMBH}}
\def\smbh#1{supermassive black hole#1 (SMBH#1)\gdef\smbh{SMBH}}

\def\bbh#1{binary black hole#1 (BBH#1)\gdef\bbh{BBH}}
\def\bh#1{black hole#1 (BH#1)\gdef\bh{BH}}
\def\ns#1{neutron star#1 (NS#1)\gdef\ns{NS}}
\def\bns#1{binary neutron star#1 (BNS#1)\gdef\bns{BNS}}
\def\nsbh#1{neutron star--black hole#1 (NSBH#1)\gdef\nsbh{NSBH}}
\def\hmns#1{hypermassive neutron star#1 (HMNS#1)\gdef\hmns{HMNS}}
\def\gw#1{gravitational wave#1 (GW#1)\gdef\gw{GW}}
\def\grb#1{Gamma-ray burst#1 (GRB#1)\gdef\grb{GRB}}
\def\pnw#1{post-Newtonian#1 (PN#1)\gdef\pnw{PN}}
\def\eos#1{equation of state#1 (EOS#1)\gdef\eos{EOS}}
\def\gpu#1{graphics processing unit#1 (GPU#1)\gdef\gpu{GPU}}
\def\sn#1{supernova#1 (SN#1)\gdef\sn{SN}}
\def\lae#1{Lyman Alpha Emitter#1 (LAE#1)\gdef\lae{LAE}}
\def\sfr#1{star formation rate#1 (SFR#1)\gdef\sfr{SFR}}
\def\sed#1{spectral energy density#1 (SED#1)\gdef\sed{SED}}
\def\lsst#1{Large Synaptic Survey Telescope#1 (LSST#1)\gdef\lsst{LSST}}
\def\rproc#1{rapid neutron capture#1 ($r$-process#1)\gdef\rproc{$r$-process#1}}

\newacronym{\grhd}{GRHD}{general relativistic hydrodynamics}
\newacronym{\grmhd}{GRMHD}{general relativistic magnetohydrodynamics}
\newacronym{\gw}{GW}{gravitational wave}
\newacronym{\EM}{EM}{electromagnetic}
\newacronym{\utb}{UTB}{University of Texas (Brownsville)}
\newacronym{\aps}{APS}{American Physical Society}
\newacronym{\lsu}{LSU}{Louisiana State University}
\newacronym{\rit}{RIT}{Rochester Institute of Technology}
\newcommand{\nn}{\nonumber}

\graphicspath{{./plots/}}

\newcommand{\macro}[1]{\textcolor{black}{#1}}
\newcommand{\hrssbestmsmag}{\macro{\ensuremath{2.7\times10^{-22}}}}
\newcommand{\hrssbestecbc}{\macro{\ensuremath{9.6\times10^{-22}}}}
\newcommand{\otwoduration}{\macro{118.3}}
\newcommand{\livetimerange}{\macro{114.7 to 118.3}}
\newcommand{\wvfnumber}{\macro{13}}
\newcommand{\far}{\macro{$1/50$}}
\newcommand{\calamp}{\macro{$7\%$}}
\newcommand{\calpha}{\macro{$3$}}
\newcommand{\inspiralB}{\macro{30 Mpc}}
\newcommand{\rateimprov}{\macro{$\sim 30\%$}}

\newtoggle{endauthorlist}
\toggletrue{endauthorlist}

\begin{document}

\title{All-sky search for long-duration gravitational-wave transients in the second Advanced LIGO observing run}


\iftoggle{endauthorlist}{
  %
  %
  \let\mymaketitle\maketitle
  \let\myauthor\author
  \let\myaffiliation\affiliation
  \author{The LIGO Scientific Collaboration}
  \author{The Virgo Collaboration}
}{
  %
  %
}

\begin{abstract}
We present the results of a search for long-duration gravitational-wave transients in the data from the Advanced LIGO second observation run; we search for gravitational-wave transients of $2~\text{--}~ 500$~s duration in the $24 - 2048$\,Hz frequency band with minimal assumptions about signal properties such as waveform morphologies, polarization, sky location or time of occurrence. 
Signal families covered by these search algorithms include fallback accretion onto neutron stars, broadband chirps from innermost stable circular orbit waves around rotating black holes, eccentric inspiral-merger-ringdown compact binary coalescence waveforms, and other models. 
The second observation run totals about \otwoduration~days of coincident data between November 2016 and August 2017. 
We find no significant events within the parameter space that we searched, apart from the already-reported binary neutron star merger GW170817.
We thus report sensitivity limits on the root-sum-square strain amplitude $h_{\mathrm{rss}}$ at $50\%$ efficiency. These sensitivity estimates are an improvement relative to the first observing run and also done with an enlarged set of gravitational-wave transient waveforms.
Overall, the best search sensitivity is $h_{\mathrm{rss}}^{50\%}$=\hrssbestmsmag~$\mathrm{Hz^{-1/2}}$ for a millisecond magnetar model. For eccentric compact binary coalescence signals, the search sensitivity reaches $h_{\mathrm{rss}}^{50\%}$=\hrssbestecbc~$\mathrm{Hz^{-1/2}}$.
\end{abstract}

\section{Introduction}

The second observation run of the Advanced LIGO~\citep{aLIGO} and Advanced Virgo~\citep{adVirgo} detectors ushered in the era of multi-messenger astronomy. In addition to the detection of further binary black hole systems~\citep{AbEA2017a,AbEA2017c,AbEA2017d}, the first binary neutron star system GW170817~\citep{AbEA2017b}, associated with GRB 170817A~\citep{GRBGWPaper} and corresponding electromagnetic radiation AT 2017gfo~\citep{MMAPaper}, were jointly detected. 
This led to searches for a post-merger signal from the binary neutron star event, including on the timescales presented in this paper~\citep{AbEA2017k,AbEA2018b}. 
In this paper, we update the results of the unmodeled long-duration transient search from the first Advanced LIGO observing run~\citep{AbEA2017g} with the data from the second observing run. 

We use four pipelines, described below, with different responses across the parameter space, providing complementary coverage of the signal models we are interested in.
The search was motivated by a wide range of poorly understood astrophysical phenomena for which predictive models are not readily available; these include fallback accretion, accretion disk instabilities and non-axisymmetric deformations in magnetars. 
Fallback accretion of ejected mass in newborn neutron stars can lead to deformation, causing the emission of gravitational waves until the star collapses into a black hole~\citep{Lai1995,PiOt2011,PiTh2012}. Accretion disk instabilities and fragmentation can cause stellar material to spiral in a black hole, emitting relatively long-lived gravitational waves~\citep{PiPf2007,VP2001,VP2008}.
Non-axisymmetric deformations in magnetars, proposed as progenitors of long and short gamma-ray bursts~\citep{MeGi2011,RoOb2013}, can also emit gravitational waves~\citep{CoMe2009}.
Moreover, we introduce new waveforms families based on astrophysical phenoma such as fallback accretion down to the innermost stable circular orbit of a rapidly rotating black hole~\citep{VP2016}, highly eccentric binary black hole coalescences~\citep{HuMo2017}, and gamma-ray burst \& X-ray events~\citep{CoMe2009}.

Although this analysis targets sources for which the gravitational waveform is not well-described, it is possible for the long-duration searches to detect low-mass compact binary coalescences, typically searched for with matched filtering techniques.
As discussed in other publications~\citep{AbEA2017b}, the data containing the gravitational-wave signal resulting from GW170817 is corrupted by the presence of a short-duration (less than 5 ms), powerful transient noise event in one of the detectors~\citep{AbEA2017b}. Using a data set where this short transient has been subtracted from the LIGO-Livingston data stream, the GW170817 signal is the most significant event of the search.
As the searches reported in this paper does not add significantly to the many other studies carried out for this event~\citep{AbEA2017b,AbEA2018d,AbEA2018e,AbEA2017k}, it has been decided to keep the original data set, veto the large transient noise and focus on any other long duration gravitational-wave signals.

The paper is organized as follows.
We describe the data used in the analysis in Section~\ref{sec:dataset}.
The algorithms used to analyze the data are outlined in Section~\ref{sec:searches}.
The results of the analysis and their implications are discussed in Section~\ref{sec:results}.
Section~\ref{sec:conclusion} provides our conclusions and avenues for future research.

\section{Data}
\label{sec:dataset}

The second observation run lasted from November 25, 2016 to August 25, 2017.
Between the first and second observing runs, a series of fixes and upgrades of the two LIGO detectors in Hanford, WA and Livingston, LA, allowed the run to begin with LIGO detectors' sensitivity reaching a binary neutron star range of $\sim$ 80 Mpc -- please see  \citet{AbEA2018c} for a discussion of the range metric. 
Thanks to commissioning break periods, Livingston's sensitivity increased steadily during the second observation run, finally reaching 100\,Mpc. 
LIGO Hanford suffered from a 5.8 magnitude earthquake in Montana on July $\mathrm{6^{th}}$~2017,
which induced a 10\,Mpc drop in sensitivity, and this was not recovered during the science run.
On August $\mathrm{1^{st}}$, the Virgo detector joined the run with a binary neutron star range of 26 Mpc. It has been shown that adding the one-month Virgo data set does not improve the search sensitivity mainly because of the sensitivity difference between the detectors. 
We thus report the results of a two LIGO detector coincident search. The overlap in time when both detectors are taking data in suitable for analysis was approximately \otwoduration~days. 
The effective coincident time analyzed by each pipeline depends on the data segmentation choice and lies in the range \livetimerange~days.

Coincident data contains a large number of non-Gaussian transient noise events (glitches) of instrumental or environmental origin that mimic the characteristic of the targeted signals. 
For the first time, well identified sources of noise have been subtracted from the LIGO data~\citep{Davis:2018yrz}. 
Yet, some glitches, typically lasting from a few milliseconds up to few seconds and varying widely in frequency, remain. Their presence, even the very short ones, may negatively impact the sensitivity of the searches~\citep{AbEA2018}. 
Time varying spectral lines are also a source of noise events for the long-duration transient searches. 
To veto these transient noise events, each pipeline implements specific glitch rejection criteria; because the search targets long-duration signals, short-duration glitches, which are usually the most problematic sources of noise, are easily suppressed. 
The next section provides more details about the noise rejection procedures that also may include data quality vetoes based on correlations with auxiliary channels~\citep{Aasi:2012wd,Abbott:2016ctn}. 

\section{Searches}
\label{sec:searches}

As in the previous analysis, we use four pipelines to search for transients that last between $2~\text{--}~ 500$~s and span a frequency band of $24~\text{--}~ 2048$~Hz.
The use of multiple pipelines provides redundancy, and due to the differences in the clustering algorithms, lead to different sensitivities to different waveform morphologies or parts of the parameter space.
Unmodeled searches for gravitational waves typically cast the analysis as pattern recognition problems.
Gravitational-wave time-series are Fourier transformed in chunks of time, and spectrograms are created based on statistics derived from these Fourier transforms.
Then pattern recognition algorithms are used to search for patterns, corresponding to gravitational waves, within spectrograms.
In general, these consist of two classes. 
The first is seed-based~\citep{KhCh2009,Pr2016}, where thresholds are placed on pixel values in the spectrograms and pixels above this threshold are clustered together.
The second is seedless~\citep{ThCo2013,ThCo2014}, where tracks are constructed from a generic model and integrated across the spectrograms; in this analysis, we use B\'ezier curves~\citep{Far1996,ThCo2013,ThCo2014,CoMe2015b,ThCo2015}.

The pipelines used are the long-duration configuration of Coherent WaveBurst (cWB)~\citep{KlVe2016}, two different versions of the Stochastic Transient Analysis Multi-detector Pipeline - all sky (STAMP-AS) pipeline~\citep{Pr2016,ThCo2015}, and the X-pipeline Spherical Radiometer (X-SphRad)~\citep{Fa2017}.
These pipelines are the same, or slightly updated versions, of those used in the search for long-duration transients in the first observation run and fully described in~\citep{AbEA2017f}.
cWB is based on a maximum-likelihood-ratio statistic, built as a sum of excess power coherent between multiple detectors in the time-frequency representation of the interferometer responses~\citep{KlVe2016}. 
The search is performed in the frequency range $24~\text{--}~ 2048 $~Hz, on data where all poor quality periods have been discarded.
The trigger events surviving the selection criteria to reject glitches are ranked according to their detection statistic $\eta_{c}$, which is related to the coherent signal-to-noise ratio (SNR). 
The selection criteria require the coherence coefficient $c_{c}$ to be larger than 0.6, and the weighted duration of the candidate to be larger than \unit[1.5]{s}. 
The first measures the degree of correlation between the detectors, while the latter measures the duration weighted by the excess power amplitude of the pixel on the time-frequency likelihood map.
The trigger events are then divided in two samples according to their estimated mean frequency: \unit[24-200]{Hz} and \unit[200-2048]{Hz}. This allows for the isolation of the unexpected higher rate of glitches at low frequency during the first half of the O2 observation run.
STAMP-AS uses the cross-correlation of data from two detectors to create coherent time-frequency maps of cross-power SNR with a pixel size of \unit[1]{s} $\times$ \unit[1]{Hz} covering $24~\text{--}~ 2000$~Hz in combination with a seed-based (Zebragard) and seedless (Lonetrack) clustering algorithm.
Significant spectral features, including wandering lines, are masked in the creation of the spectrograms. 
As in the search during the first observing run, Zebragard eliminates the short duration glitches by requesting that the fraction of SNR in each time bin be smaller than 0.5 and that the SNR ratio between the two detectors be smaller than 3. 
The X-pipeline Spherical Radiometer (X-SphRad) uses an X-pipeline~\citep{SuJo2010} back end in combination with a fast cross-correlator in the spherical harmonic domain~\citep{Can2007} to search for gravitational-wave transients in the $24~\text{--}~ 1000$~Hz frequency range. The method allows for the data to be processed independently of sky position and avoids redundant computations. A next-nearest-neighbor clustering algorithm is applied on a time-frequency representation of the data with a resolution of \unit[1]{s} $\times$ \unit[1]{Hz} to form trigger events, which are then ranked by the ratio of the sum of power in all the l$>$0 spherical harmonic modes to that in the l$=$0 mode. 
Significant spectral features such as standing power lines are removed using a zero-phase linear predictor filter that estimates the power spectrum and whitens the data~\citep{chatterji2005}.
Finally, X-SphRad eliminates triggers that coincide with poor quality data periods that have been identified using auxiliary channels. These periods are excluded from the analysis time by cWB, and STAMP-AS Zebragard analysis selects a subset of them according to a procedure described in~\citep{Fr2018}.

The false alarm rate of each search is estimated as a function of the pipeline's ranking statistic. Each use the data to perform this estimate, as opposed to a Gaussian approximation, because of the significant non-Gaussianity of the data, transient noise, and the non-stationarity of some of the spectral features.
These glitches have a variety of causes, both environmentally driven such as from seismic events~\citep{MaFa2012,CoEa2017} or magnetic fields~\citep{KoBi2017,CoCh2016}, and instrumental effects, such as test mass suspension glitches~\citep{WaAb2017} and other sources of spectral features~\citep{CoLi2010}.
For all of the pipelines in this analysis, the correlation of data in different detectors is used to exclude data transients which are unlikely to be of astrophysical origin.
To estimate the background for all pipelines used in this analysis, the time-slides methodology is applied~\citep{WaBi2010a,WaBi2010b}, each one implementing its own version.
The fundamental idea is to shift the detector data with non-physical relative time delays to eliminate any correlation from gravitational waves and re-analyze the data.
The procedure is repeated until a total of 50 years coincident detector time has been analyzed, allowing us to estimate false alarm rates at the level of 1 event in 50 years.
\section{Results}
\label{sec:results}

\begin{figure}[hbtp!]  
\centering  
\includegraphics[width=3.5in]{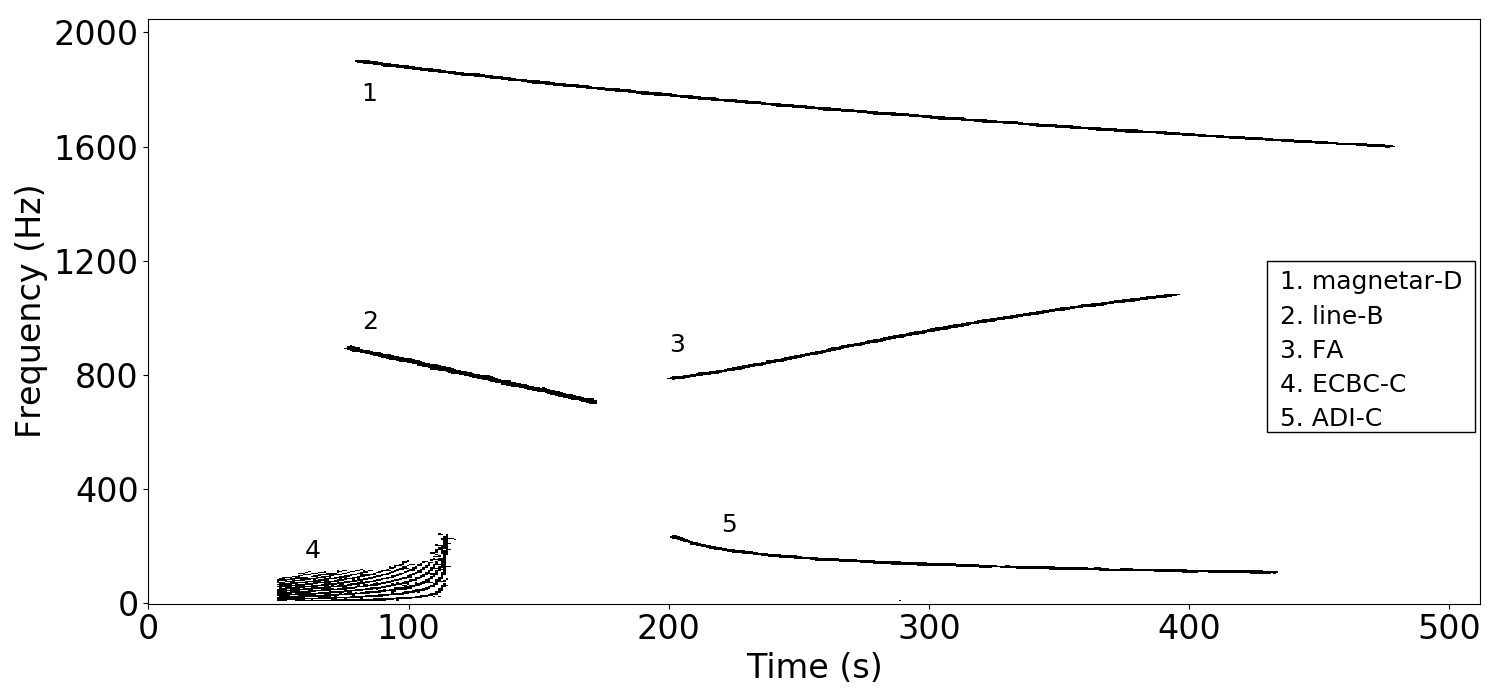}  
\caption{Time-frequency representations of a few model signals used in the search, showing a mix of chirp-up (FA, ECBC) \& chirp-down (Magnetar, ADI) astrophysical waveforms as well as a linearly decreasing ad-hoc waveform (LINE). Descriptions of these waveforms and others are given in Section~\ref{sec:results}. The harmonics of ECBC are also visible. The full set of waveforms ($\sim$ 70) chosen for this analysis fully covers the search frequency band of 24-2000\,Hz. The waveforms are shifted in time to show how they cover the parameter space in this axis as well.}  
\label{fig:O2_coverage_parameter_space}
\end{figure}

\begin{table}[h]
\begin{tabular}{ccccc}
\hline
\hline
Pipeline   & FAR   & p-value & Frequency  & Duration\\
           & [Hz]  &     & [Hz]       & [s]\\
\hline
\hline
cWB        & $1.4 \times 10^{-7}$   & 0.75 & 53--69 & 11\\

Zebragard  & $2.5 \times 10^{-7}$   & 0.92 & 1649--1753 & 29\\

Lonetrack  & $7.9 \times 10^{-8}$   & 0.80 & 608--1344 & 463\\

X-SphRad   & $9.7 \times 10^{-8}$  & 0.60 & 435-443   & 3\\
\hline
\hline
\end{tabular}
\caption{Properties of the most significant coincident triggers found by each of the long-duration transient search pipelines during the second observation run. FAR stands for false alarm rate, while the p-value is the probability of observing at least 1 noise trigger at higher significance than the most significant coincident trigger.}
\label{tab1}
\end{table}

None of the pipelines find a significant excess of coincident events. The most significant events found by each pipeline are reported in Table~\ref{tab1}. Their false alarm rate is in agreement with the expected background estimation.
Given the absence of a detection, we can derive upper limits on long-duration gravitational-wave transients' strain amplitude. 
A usual measure of gravitational-wave amplitude is the root-sum-square strain amplitude at the Earth, $h_{\mathrm{rss}}$,
\begin{equation}\label{eq:hrss}
h_{\mathrm{rss}} = \sqrt{\int_{-\infty}^{\infty} \left(h^2_+(t) + h^2_{\times}(t) \right) \diff t},
\end{equation}
where $h_{+}$ and $h_{\times}$ are signal polarizations at Earth's center expressed in the source frame.
We can relate this quantity to the gravitational-wave energy radiated by a source emitting isotropically at a given central frequency $f_0$~\citep{2013arXiv1304.0210S}
\begin{align}\label{eq:energy}
E_{\mathrm{gw}}^{\mathrm{iso}} & =  \frac{\pi c^3}{G} \dist^2 \int \mathrm{d}f~f^2 \left(|\tilde{h}_+(f)|^2 + |\tilde{h}_{\times}(f)|^2 \right) \nn \\
 & \approx \frac{\pi^2 c^3}{G} \dist^2 f_0^2 h_{\mathrm{rss}}^2,
\end{align}
where $\dist$ is the distance to the source and $\tilde{h}$ indicates a Fourier transform.
To estimate the $h_{\mathrm{rss}}$ at 50\% detection efficiency, we add simulated waveforms coherently to detector data, uniformly distributed in time and over sky locations.  
The waveform polarization angle and the cosine of the inclination are also varied uniformly.
Waveforms are generated at a variety of distances (or equivalently $h_{\mathrm{rss}}$) such that the 50\% detection efficiency is well-measured.
The events reconstructed are then ``detected'' if their false alarm rate is lower than the chosen value of \far~years. 

We use \wvfnumber~ families of simulated gravitational-wave signals to estimate the sensitivity of each pipeline.
The waveform families include a variety of astrophysically motivated waveforms and ad-hoc waveform models. For the astrophysical models, we include fallback accretion onto neutron stars (FA)~\citep{PiTh2012}, broadband chirps from innermost stable circular orbit waves around rotating black holes (ISCOchirp)~\citep{VP2016}, inspiral-only compact binary coalescence waveforms up to 2nd post-Newtonian order~\citep{BlIy1996} (CBC), eccentric inspiral-merger-ringdown compact binary coalescence waveforms (ECBC)~\citep{HuMo2017}, secular bar-mode instabilities in post-merger remnants~\citep{Lai1995,CoMe2009}, newly formed magnetar powering a gamma-ray burst plateau (GRBplateau)~\citep{CoMe2009}, black hole accretion disk instabilities (ADI)~\citep{VP2001}, post-merger magnetars (magnetar)~\citep{DaGi2015}, and neutron star spin down waveforms (MSmagnetar)~\citep{LaLe2017,SaLa2018}. For the ad-hoc waveforms, we include monochromatic waveforms (MONO), waveforms with a linear (LINE), quadratic (QUAD) frequency evolution, white noise band-limited (WNB) and sine-Gaussian bursts (SG). 
The waveforms are designed to span a range of astrophysical models, as well as a wide duration and frequency parameter space to test the response of the algorithms across the parameter space. Figure \ref{fig:O2_coverage_parameter_space} shows the coverage of a representative sample of the simulation set in the time-frequency space. The frequency band 10-300\,Hz are well covered with the GRBplateau and ADI families. Astrophysical waveform families such as ISCOchirp and magnetar are characterised by a wide frequency coverage and populate the higher frequency band 700-2000\,Hz. Ad-hoc waveforms families such as MONO, LINE, QUAD, WNB and SG span a wide frequency range and covers the band 50-800\,Hz, filling in any potential gap in coverage from the other models. 

\begin{figure*}[hbtp!]
  \centering
  \includegraphics[width=3.5in]{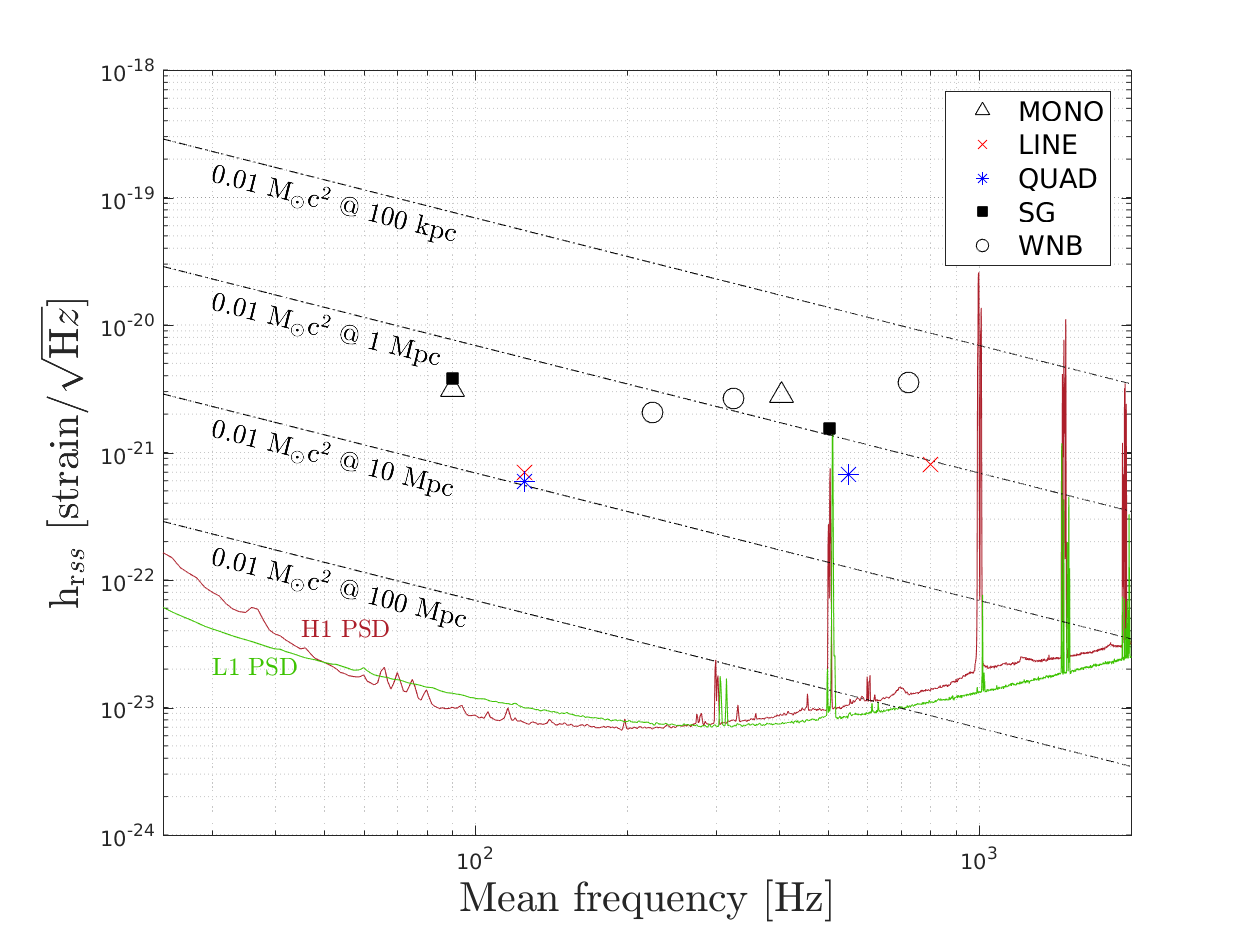}
  \includegraphics[width=3.5in]{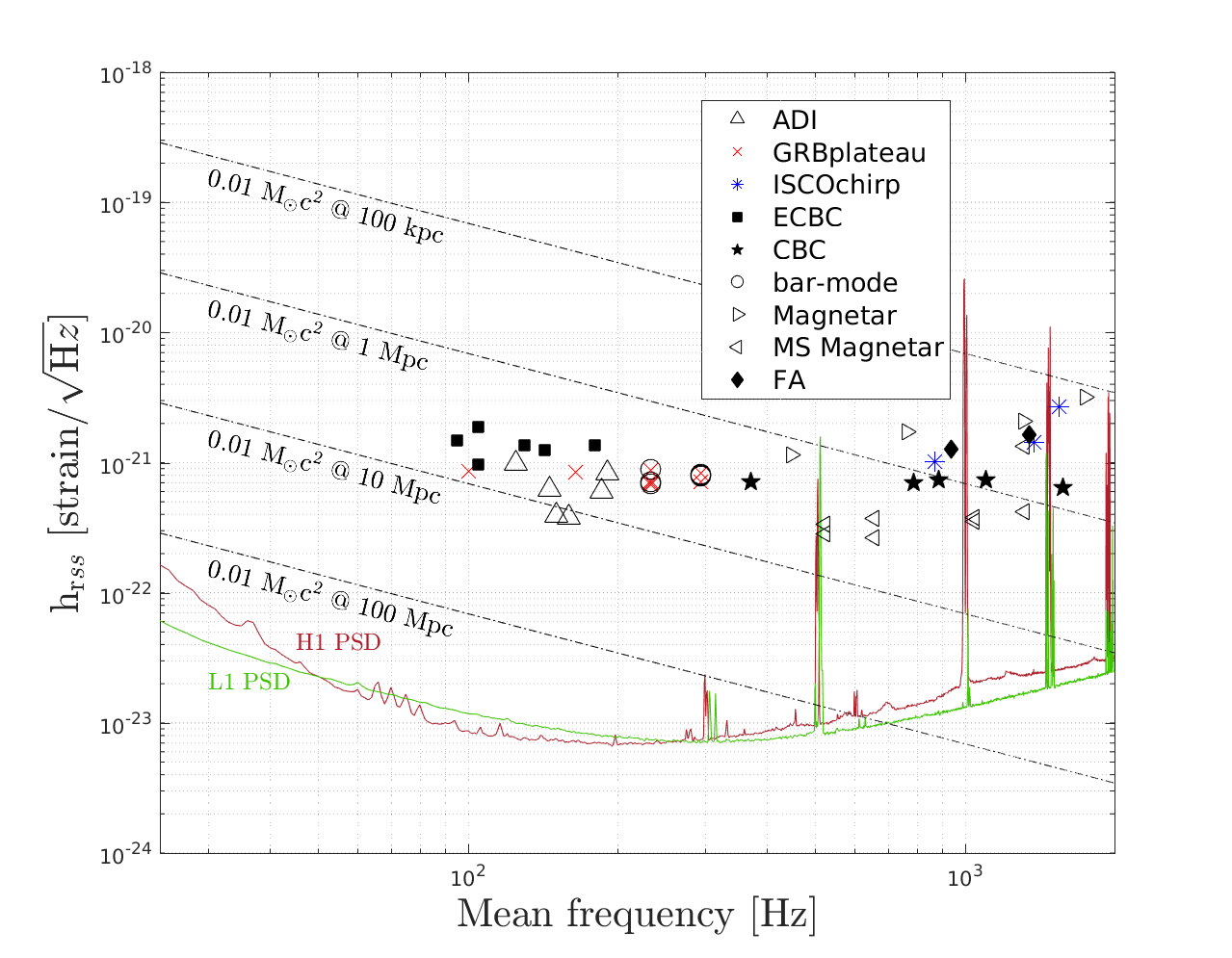}  
  \caption{Upper limits on gravitational-wave strain versus frequency for sources detected with 50\% efficiency and a false alarm rate of 1 event in 50 years. The lowest value among all 4 pipelines is represented on the plots. The left figure shows the ``ad-hoc'' waveforms' results while the ``physical'' waveforms are represented on the right. The average amplitude spectral density curves for both Hanford and Livingston are also shown.}
  \label{fig:upperlimits}
\end{figure*}

\begin{figure}[hbtp!]
  \centering
  \includegraphics[width=3.55in]{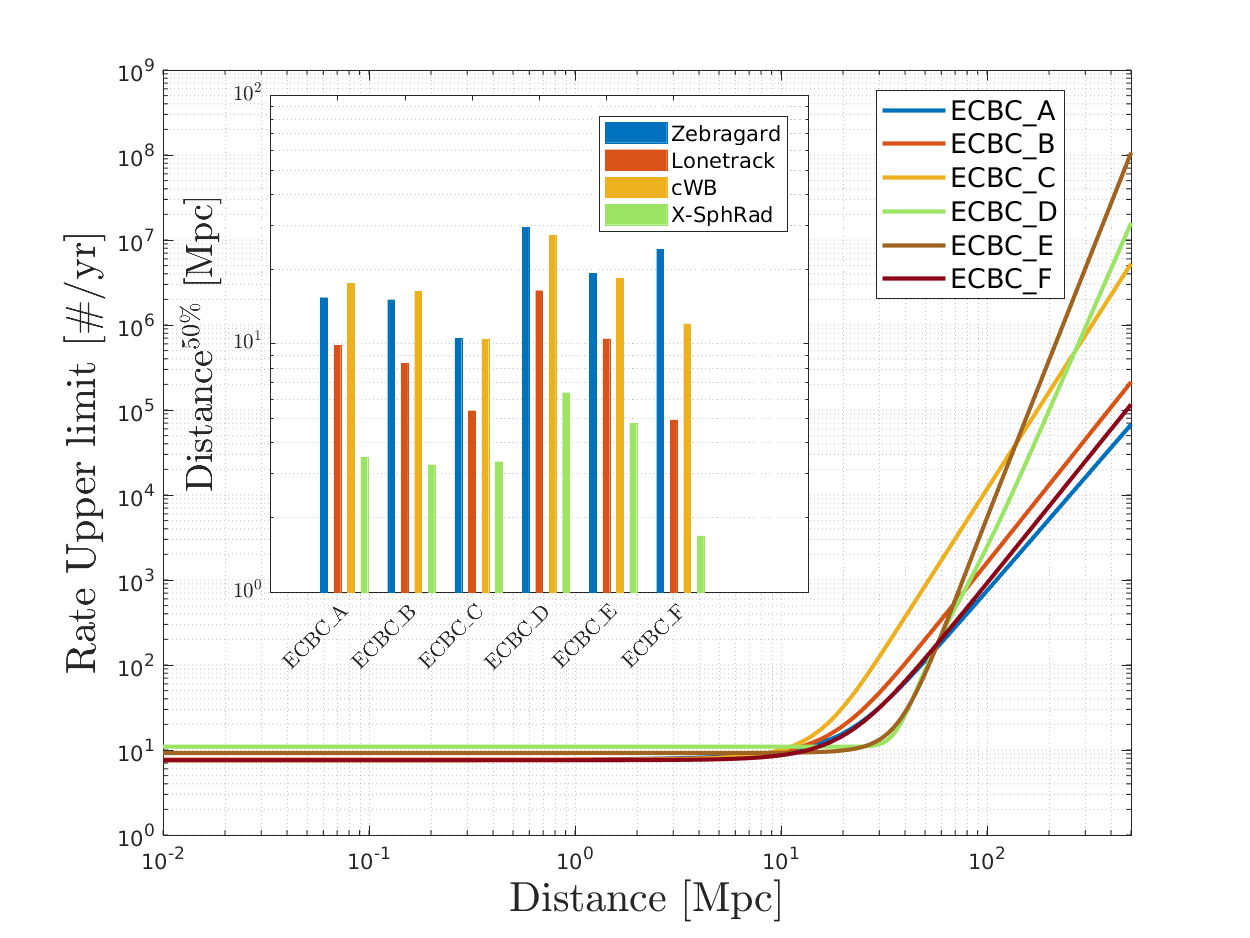}
  \caption{Upper limits (marginalizing over the second observation run amplitude calibration errors) on eccentric compact binary coalescences as a function of the distance at a 90\% confidence level considering the best results for each waveform. The inset shows the distance at 50\% detection efficiency for the pipelines in this analysis for comparison. ECBC\_A, ECBC\_B, and ECBC\_C are $1.4-1.4$ solar mass binaries with eccentricities of 0.2, 0.4, and 0.6 respectively, while ECBC\_D, ECBC\_E, and ECBC\_F are $3.0-3.0$ solar mass binaries with eccentricities of 0.2, 0.4, and 0.6 respectively, where the masses are quoted in the detector frame.}
  \label{fig:rate}
\end{figure}

In Figure~\ref{fig:upperlimits}, we show the best results among all pipelines for almost all waveforms.
We also compute the 90\% confidence level limit on the rate of long-duration gravitational-wave transients assuming a Poissonian distribution of sources. To do so, we use the loudest event statistic method~\citep{BrCr2004}. We fold in the systematic uncertainty that arises from the strain amplitude calibration, which is \calamp~ in amplitude and \calpha~ degrees in phase, a conservative number used for both instruments in the frequency band analyzed here~\citep{CaBe2017}.

Figure~\ref{fig:rate} shows the rate as a function of distance for the eccentric compact binary coalescence signals considered in this analysis.
For a $1.4-1.4$ solar mass binary with an eccentricity of 0.4, the 50\% efficiency distance is \inspiralB. For comparison, this is more than a factor 2 lower than what matched filter searches could reach for $1.4-1.4$ solar mass binaries with no eccentricity during the second observation run~\citep{AbEA2017b}.
Due to the improved sensitivity and greater duration of the second observation run above and beyond the first observation run, the rate limits for models used in previous analyses improved by a factor of \rateimprov.
The detection distances vary significantly from one signal to another.
For example, the ADI waveforms have distance limits of tens of megaparsecs, while the magnetar waveforms have limits of tens of kiloparsecs.
The difference in ranges is due mainly to the energy budget of the system, but also due to the overall signal morphologies, which can be more or less difficult for the pipeline clustering techniques to recover entirely.

\section{Conclusions}
\label{sec:conclusion}

We have performed an all-sky search for unmodeled long-duration gravitational-wave transients in the second observing run. This search did not lead to the detection of any new gravitational waves.
In addition to the intrinsic gain due to detectors' sensitivity improvement and the length of the observing run, we have increased significantly the number of waveforms used to estimate the pipelines' sensitivity.
The theoretical uncertainties of the models used are rather large, including the mechanisms, their amplitudes, and their potential rates, although it is likely we are sensitive to relatively small amplitude emissions within the Local Group.

With the recent arrival of Advanced Virgo to the advanced gravitational-wave detector network, its future improvements will merit its inclusion in analyses in the next observing runs. Overall, the expectation is that the design sensitivities for the gravitational-wave networks will yield gains of up to a factor of 10, depending on the frequency range considered \citep{AbEA2018c}.

\acknowledgements
The authors gratefully acknowledge the support of the United States
National Science Foundation (NSF) for the construction and operation of the
LIGO Laboratory and Advanced LIGO as well as the Science and Technology Facilities Council (STFC) of the
United Kingdom, the Max-Planck-Society (MPS), and the State of
Niedersachsen/Germany for support of the construction of Advanced LIGO 
and construction and operation of the GEO600 detector. 
Additional support for Advanced LIGO was provided by the Australian Research Council.
The authors gratefully acknowledge the Italian Istituto Nazionale di Fisica Nucleare (INFN),  
the French Centre National de la Recherche Scientifique (CNRS) and
the Foundation for Fundamental Research on Matter supported by the Netherlands Organisation for Scientific Research, 
for the construction and operation of the Virgo detector
and the creation and support  of the EGO consortium. 
The authors also gratefully acknowledge research support from these agencies as well as by 
the Council of Scientific and Industrial Research of India, 
the Department of Science and Technology, India,
the Science \& Engineering Research Board (SERB), India,
the Ministry of Human Resource Development, India,
the Spanish  Agencia Estatal de Investigaci\'on,
the  Vicepresid\`encia i Conselleria d'Innovaci\'o, Recerca i Turisme and the Conselleria d'Educaci\'o i Universitat del Govern de les Illes Balears,
the Conselleria d'Educaci\'o, Investigaci\'o, Cultura i Esport de la Generalitat Valenciana,
the National Science Centre of Poland,
the Swiss National Science Foundation (SNSF),
the Russian Foundation for Basic Research, 
the Russian Science Foundation,
the European Commission,
the European Regional Development Funds (ERDF),
the Royal Society, 
the Scottish Funding Council, 
the Scottish Universities Physics Alliance, 
the Hungarian Scientific Research Fund (OTKA),
the Lyon Institute of Origins (LIO),
the National Research, Development and Innovation Office Hungary (NKFI), 
the National Research Foundation of Korea,
Industry Canada and the Province of Ontario through the Ministry of Economic Development and Innovation, 
the Natural Science and Engineering Research Council Canada,
the Canadian Institute for Advanced Research,
the Brazilian Ministry of Science, Technology, Innovations, and Communications,
the International Center for Theoretical Physics South American Institute for Fundamental Research (ICTP-SAIFR), 
the Research Grants Council of Hong Kong,
the National Natural Science Foundation of China (NSFC),
the Leverhulme Trust, 
the Research Corporation, 
the Ministry of Science and Technology (MOST), Taiwan
and
the Kavli Foundation.
The authors gratefully acknowledge the support of the NSF, STFC, MPS, INFN, CNRS and the
State of Niedersachsen/Germany for provision of computational resources.

\bibliographystyle{aasjournal}
\bibliography{references}

\iftoggle{endauthorlist}{
  %
  %
  \let\author\myauthor
  \let\affiliation\myaffiliation
  \let\maketitle\mymaketitle
  \title{Authors}
  \author{B.~P.~Abbott}
\affiliation{LIGO, California Institute of Technology, Pasadena, CA 91125, USA}
\author{R.~Abbott}
\affiliation{LIGO, California Institute of Technology, Pasadena, CA 91125, USA}
\author{T.~D.~Abbott}
\affiliation{Louisiana State University, Baton Rouge, LA 70803, USA}
\author{S.~Abraham}
\affiliation{Inter-University Centre for Astronomy and Astrophysics, Pune 411007, India}
\author{F.~Acernese}
\affiliation{Dipartimento di Farmacia, Universit\`a di Salerno, I-84084 Fisciano, Salerno, Italy}
\affiliation{INFN, Sezione di Napoli, Complesso Universitario di Monte S.Angelo, I-80126 Napoli, Italy}
\author{K.~Ackley}
\affiliation{OzGrav, School of Physics \& Astronomy, Monash University, Clayton 3800, Victoria, Australia}
\author{C.~Adams}
\affiliation{LIGO Livingston Observatory, Livingston, LA 70754, USA}
\author{R.~X.~Adhikari}
\affiliation{LIGO, California Institute of Technology, Pasadena, CA 91125, USA}
\author{V.~B.~Adya}
\affiliation{OzGrav, Australian National University, Canberra, Australian Capital Territory 0200, Australia}
\author{C.~Affeldt}
\affiliation{Max Planck Institute for Gravitational Physics (Albert Einstein Institute), D-30167 Hannover, Germany}
\affiliation{Leibniz Universit\"at Hannover, D-30167 Hannover, Germany}
% ...

  \newpage
  \maketitle
}

\end{document}